# CFM6, a closed-form NLI EGN model supporting multiband transmission with arbitrary Raman amplification

Yanchao Jiang, Pierluigi Poggiolini

OptCom, Politecnico di Torino, Italy

**Abstract:** We formulated a closed-form EGN model for nonlinear interference in ultra-wideband optical systems with arbitrary Raman amplification. This model enhanced the CISCO-POLITO-CFM5 performance by introducing a novel contribution attributed to the backward Raman amplification. It can handle the frequency-dependent fiber parameters and inter-channel stimulated Raman scattering.

## 1. Introduction

In the realm of long-haul optical systems, numerous technologies are currently competing in the quest for increasing the throughput. They can be broadly classified into two categories: space-division multiplexing (SDM) and ultra-wideband (UWB). All SDM and some UWB technologies require that new cables be deployed. A notable exception is UWB over existing standard single-mode fiber (SMF) cables, a potentially attractive alternative for carriers who do not wish to lay out new cables.

UWB over SMF consists of extending the transmission bandwidth beyond the C band. The first step, the extension to L-band, is already commercially available. Research is now focusing on other bands, primarily S and O but also E and U. In the context of long-haul systems, it is mostly the S-band that is being considered, because higher frequency bands such as E and O suffer from more serious propagation impairments, while the U-band appears problematic due to bend loss and non-mature amplification solutions.

Many research experiments of C+L+S transmission have already been successfully carried out [1-6]. For instance [6] where, remarkably, 12,345 km were reached, using about 3 THz of S-band (plus C and L band with 6 THz each) over special low-loss 4-core MCF and Raman amplification with 8 pumps.

To achieve commercial attractiveness of C+L+S systems in conventional terrestrial long-haul, some goals must conceivably be met: (a) the addition of the S-band must bring about a substantial throughput increase, such as 4x-5x vs. standard (4.5-THz) C-band-only systems; (b) the operating conditions in the three bands should be rather uniform (similar GSNRs); (c) if used, Raman amplification must need a small number of limited power pumps.

To pursue these goals, complex joint optimization needs to be carried out of link and system parameters, as well as WDM launch power profiles and Raman pump frequencies and powers. This requires fast and accurate physical layer models, capable of accounting for the broadband-dependence of all key fiber and system parameters, together with Inter-channel Raman Scattering (ISRS) and Raman amplification. UWB Closed-Form Models (CFMs) have been developed for this purpose: two groups, one at UCL, and one at PoliTo (in collaboration with CISCO), have independently obtained UWB CFMs based on approximations of the GN/EGN models, with similar foundations but with differences in features and final analytical form. For the UCL CFM see [7-10], for the CISCO-PoliTo CFM see [11-14].

In this paper, we focus on the derivation of CFM6 with backward Raman amplification. It upgrades the CISCO-POLITO-CFM5 ([12]) and introduces a new contribution for backward Raman amplification. We start from the loss model to approximate the actual power profiles. Then, we provide the details on the derivation.

## 2. The loss model

To calculate the nonlinear interference (NLI) fully in closed form, we need to find out an analytical model to represent the channel loss coefficient over a given span. Within the framework of CFM5, three distinct parameters are employed to characterize the filed loss in the $i$-th channel:

$$\alpha_i(z) = \alpha_{0,i} + \alpha_{1,i} \cdot \exp(-\sigma_i \cdot z) \tag{1}$$

The model exhibits robust performance in accounting for the influence of the ISRS effect and forward Raman amplification. However, it fails in addressing the impact of backward Raman amplification. This limitation becomes particularly pronounced when a significantly wide band is required, where incorporating backward Raman amplification becomes essential to counteract power losses in the high-frequency channels due to the strong ISRS effect.

We start from the coupled differential Raman equations in two directions for arbitrary Raman amplifications, and the ISRS effect. The power evolution within the $i$-th channel is modeled as,

$$\pm \frac{dP_i(z)}{dz} = \left\{ \sum_{j=1}^{N_{ch}+N_p} \varsigma\left(\frac{f_i}{f_j}\right) \times C_R(f_j - f_i) \times P_j(z) \right\} \times P_i(z) - 2 \times \alpha_{i,\text{lin}} \times P_i(z) \tag{2}$$

The symbols '$\pm$' represent the two directions of propagation, where '+' denotes forward-propagating signals or pumps, and '-' denotes backward-propagating pumps. There are $N_{ch}$ signal channels and $N_p$ pump channels. The $i$-th channel is characterized by its center frequency, denoted as $f_i$, and $f_i < f_j, 1 \leq i < j \leq N_{ch} + N_p$.

$\varsigma(f_i/f_j)$ can be simplified in each pair of channels $(i, j)$ according to Eq. (3):

$$\varsigma\left(\frac{f_i}{f_j}\right) = \begin{cases} \frac{f_i}{f_j}, f_i > f_j \\ 0, f_i = f_j \\ 1, f_i < f_j \end{cases} \tag{3}$$

$C_R(f_j - f_i)$ represents the ISRS gain coefficient, characterized as an odd function with respect to the frequency spacing between each pair of channels $(i, j)$,

$$\begin{aligned} C_R(f_j - f_i) &= -C_R(f_i - f_j) \\ C_R(f_j - f_i) &\geq 0, f_i < f_j \end{aligned} \tag{4}$$

$\alpha_{i,\text{lin}}$ represents the intrinsic fiber loss without ISRS effect.

These equations could be solved numerically with excellent accuracy but could not be solved analytically in realistic scenarios. Therefore, the pursuit of an analytical expression to approximate the numerical solution as accurate as possible is crucial. This is the center to the CFM6.

Backward Raman pumps inject power at the end of the fiber. They are generally quite powerful to generate much NLI due to the amplification onto the signals. The impact on the signal power is depicted in Fig. 1, which is taken from the case study below. The signal initially experiences a power decrease due to fiber attenuation, followed by a fast power increase from the five powerful pumps.

Case-study:

- 95 km span, 3dBm flat input
- 100GBuad, frequency spacing 125GHz, roll-off 0.1
- 38 channels in C band, 38 channels in L band. The lowest and highest frequencies are shown in Fig. 2
- Fiber characterized for loss, dispersion, and nonlinearity as in [13]
- ISRS is characterized as in [13]
- 5 pumps with power of [360, 320, 200, 130, 180] mW at [210.5, 208.9, 206.7, 204.5, 200.5] THz

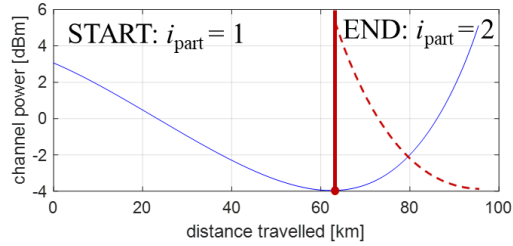

Fig. 1: The lowest-frequency L-band channel (186.1THz) power evolution along a 95-km fiber experiencing a powerful backward Raman amplification.

The idea underlying Eq. (1) is to characterize the loss using three parameters, primarily focusing on scenarios where the intrinsic fiber loss prevails. This makes it fail in accurately approximating the power profile in the presence of backward Raman pumps. It would underestimate the total NLI calculation. Despite of its limitations, Eq. (1) remains useful for modeling the initial phase of the power profile due to dominant intrinsic fiber loss. To address the challenges posed by the presence of backward Raman pumps and the subsequent deviation, an alternative loss model can be used to approximate the power profile towards the end of the fiber. This hybrid approach enables a more comprehensive representation of the power profile, enhancing the accuracy of NLI calculation.

Accordingly, each span is divided into 2 segments at the points of the lowest power (the red dot in Fig. 1):

- $i_{\text{part}} = 1$: representing the initial segment, characterized by a power profile where losses dominate over gains. The loss can be effectively modelled using Eq. (1). For clarity and differentiation from the loss model in the subsequent segment, it is re-written with the subscript 'st',

$$\alpha_{\text{st}}^{(n_s)}(z) = \alpha_{0,\text{st}}^{(n_s)} + \alpha_{1,\text{st}}^{(n_s)} \cdot \exp\left(-\sigma_{\text{st}}^{(n_s)} \cdot z\right) \qquad (5)$$

The starting point is at $z = 0$, transmitting to $L_{s,\text{st}}^{(n_s)}$. The power at any distance $z$ within this segment can be computed as,

$$\begin{aligned} P_{\text{st}}^{(n_s)}(z) &= P_{\text{st}}^{(n_s)}(0) \cdot \exp\left(-2\int_0^z \alpha_{\text{st}}^{(n_s)}(z')dz'\right), 0 \le z \le L_{s,\text{st}}^{(n_s)} \\ &= P_{\text{st}}^{(n_s)}(0) \cdot \exp\left(-2\alpha_{0,\text{st}}^{(n_s)} z + 2\alpha_{1,\text{st}}^{(n_s)} \cdot \frac{\exp\left(-\sigma_{\text{st}}^{(n_s)} \cdot z\right)-1}{\sigma_{\text{st}}^{(n_s)}}\right) \end{aligned} \qquad (6)$$

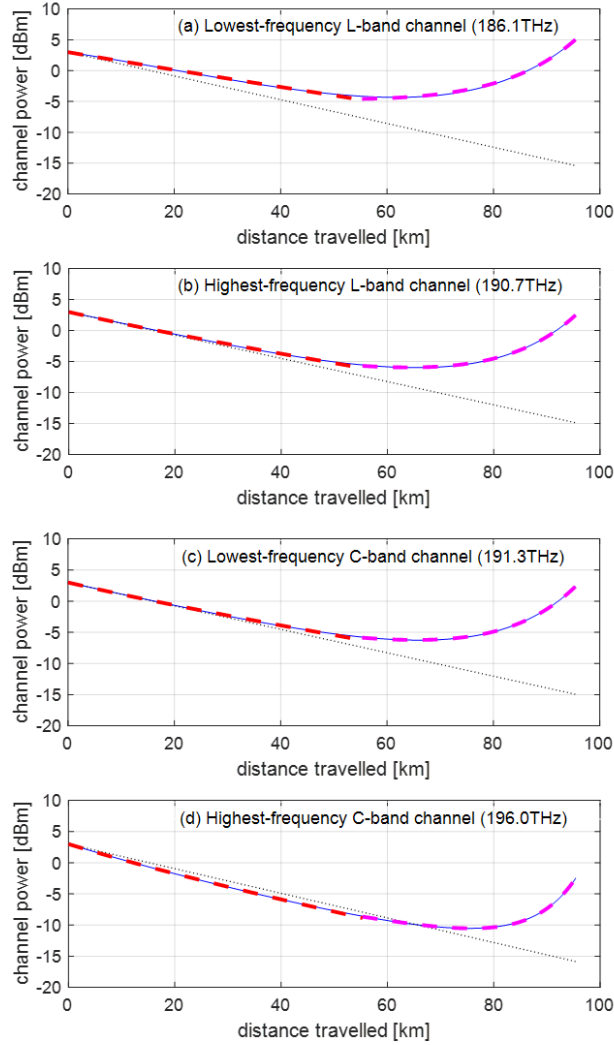

Fig. 2 Power vs. distance into a single span for the lowest and highest frequency L-band and C-band channels. Solid lines: numerical power evolution in the presence of ISRS and backward Raman amplification. Dotted lines: power evolution in the absence of ISRS. Dashed lines: power evolution assuming the loss models of Eq. (5) and Eq. (7), corresponding to the two segments in each span.

- $i_{\text{part}} = 2$: representing the latter segment, characterized by a power profile where gains from the backward pumps dominate over losses. If this segment is mirrored from left to right, as depicted by the red dashed curve in Fig. 1, the loss can also be modelled using Eq. (1),

$$\alpha_{\text{end,flip}}^{(n_s)}(z) = \alpha_{0,\text{end,flip}}^{(n_s)} + \alpha_{1,\text{end,flip}}^{(n_s)} \cdot \exp\left(-\sigma_{\text{end,flip}}^{(n_s)} \cdot z\right) \tag{7}$$

Where the subscript 'end' is used vs. 'st'. If it is treated as an independent span, spanning from 0 to $L_{s,\text{end}}^{(n_s)} = L_s^{(n_s)} - L_{s,\text{st}}^{(n_s)}$, the power at any distance $z$ within this segment can be computed as,

$$P_{\text{end-flip}}^{(n_s)}(z) \approx P_{\text{end-flip}}^{(n_s)}(0) \cdot \exp\left(-2\int_0^z \alpha_{\text{end,flip}}^{(n_s)}(z')dz'\right), 0 \leq z \leq L_{s,\text{end}}^{(n_s)}$$

$$= P_{\text{end-flip}}^{(n_s)}(0) \cdot \exp\left(-2\alpha_{0,\text{end,flip}}^{(n_s)} z + 2\alpha_{1,\text{end,flip}}^{(n_s)} \cdot \frac{\exp\left(-\sigma_{\text{end,flip}}^{(n_s)} \cdot z\right) - 1}{\sigma_{\text{end,flip}}^{(n_s)}}\right) \quad (8)$$

The parameters in both Eqs. (5) and (7) can be found using the same method as in CFM5 (Eq. (30.1) and (30.2) in [11]). The formulas minimize the Mean Square Error (MSE) between the power profile generated by Eqs. (6) and (8) and the exact power evolution. The MSE is calculated so that the error weighs more where the signal power is larger, to enhance accuracy at high power where most of the NLI is produced.

Fig.2 illustrates the different power profiles of either the lowest-frequency or highest-frequency channel across the two bands in case-study. The blue solid curves represent the actual power profile by solving Eq. (2) numerically. The dashed curves represent the best-fitting power profiles using the loss model, with the red curves obtained through Eq. (6) and the magenta curves obtained through Eq. (8). Notably, all fitted curves closely align with the actual power profiles. Additionally, the black dotted curves represent the power profiles in the absence of ISRS effect, indicating the power only experiences the intrinsic fiber loss.

## 3. CFM6

We start the derivation of CFM6 with the incoherent Gaussian noise (IGN) model. We firstly focus on the first segment within a span, where the NLI calculation is obtained using the same procedure in CFM5. We reiterate some key steps in this paper (more details see [11]). We then delve into a detailed derivation for the second segment where the backward Raman pumps jump in. This is the novel contribution in CFM6. Finally, we integrate the machine-learning factors into CFM6 to accomplish it as an approximating tool to the EGN model.

The power spectral density (PSD) of NLI based on IGN model is expressed as:

$$G_{\text{NLI}}(f) = \frac{16}{27} \sum_{m_{\text{ch}}=1}^{N_{\text{ch}}} \sum_{n_{\text{ch}}=1}^{N_{\text{ch}}} \sum_{k_{\text{ch}}=1}^{N_{\text{ch}}} G_{m_{\text{ch}}} G_{n_{\text{ch}}} G_{k_{\text{ch}}} \iint_{S(m_{\text{ch}},n_{\text{ch}},k_{\text{ch}})} |\text{LK}(f_1, f_2, f_1+f_2-f)|^2 df_1 df_2 \quad (9)$$

$$S(m_{\text{ch}}, n_{\text{ch}}, k_{\text{ch}}) = \left\{(m_{\text{ch}}, n_{\text{ch}}, k_{\text{ch}}) | 1 \leq m_{\text{ch}} \leq N_{\text{ch}}, 1 \leq n_{\text{ch}} \leq N_{\text{ch}}, 1 \leq k_{\text{ch}} \leq N_{\text{ch}}\right\}$$

where $G_{m_{\text{ch}}} / G_{n_{\text{ch}}} / G_{k_{\text{ch}}}$ are the PSD of the WDM signals launched into the fiber. Generally, it is raised-cosine with nonzero roll-off. In Eq. (9). we approximately replace it with a rectangular channel with the same center frequency as the original raised cosine channel, and we assume that the null-to-null bandwidth of the rectangular channel is equal to the symbol rate of the original channel. $\text{LK}(f_1, f_2, f_1+f_2-f)$ is the link function, which encapsulates all the parameters along the link and is explicitly given in Eq. (11).

UWB over existing SMF cables is potentially attractive. The high dispersion of SMF makes all contributions from the off-axis islands in the integration domain of Eq. (9) negligible. Therefore, we can restrict the islands to SPM and XPM to further simplify the IGN model as,

$$G_{\text{NLI}}(f) \approx \frac{16}{27} \sum_{m_{\text{ch}}=1}^{N_{\text{ch}}} G_{\text{CUT}} G_{m_{\text{ch}}}^2 \left(2 - \delta_{\text{CUT},m_{\text{ch}}}\right) \iint_{S(m_{\text{ch}},m_{\text{ch}},\text{CUT})} |\text{LK}(f_1, f_2, f_1+f_2-f)|^2 df_1 df_2 \quad (10)$$

When $m_{ch} = \text{CUT}, \delta_{\text{CUT},m_{ch}} = 1$ represents the SPM contribution. Otherwise, $\delta_{\text{CUT},m_{ch}} = 0$ represents the XPM contribution. The link function is expressed as,

$$\text{LK}(f_1, f_2, f_3) = -j\sum_{n_s=1}^{N_s} \gamma^{(n_s)}(f_1, f_2) \cdot \left\{ \prod_{p=n_s}^{N_s} \sqrt{\Gamma^{(p)}(f_1+f_2-f_3)} e^{\int_0^{L_s^{(p)}} \kappa^{(p)}(z, f_1+f_2-f_3) dz} \right\}$$
$$\cdot \left\{ \prod_{p=1}^{n_s-1} \sqrt{\Gamma^{(p)}(f_1)\Gamma^{(p)}(f_2)\Gamma^{(p)}(f_3)} e^{\int_0^{L_s^{(p)}} \left[\kappa^{(p)}(z,f_1)+\kappa^{(p)}(z,f_2)+\kappa^{*(p)}(z,f_3)\right] dz} \right\} \quad (11)$$
$$\cdot \left\{ \int_0^{L_s^{(n_s)}} e^{\int_0^{z'} \left[\kappa^{(n_s)}(z'',f_1)+\kappa^{(n_s)}(z'',f_2)+\kappa^{*(n_s)}(z'',f_3)-\kappa^{(n_s)}(z'',f_1+f_2-f_3)\right] dz''} dz' \right\}$$

Where $\gamma^{(n_s)}(f_1, f_2)$ is the fiber nonlinear coefficient and is frequency dependent,

$$\gamma^{(n_s)}(f_1, f_2) = \frac{2\pi f_1}{c} \cdot \frac{2n_2}{A_{\text{eff}}(f_1) + A_{\text{eff}}(f_2)} \quad (12)$$

$\Gamma^{(n_s)}(f_i)$ is the power gain on the $i$-th channel of the $n_s$-th span. $\kappa^{(n_s)}(z'', f_i)$ contains the linear effect in the $n_s$-th span, and is expressed as,

$$\kappa^{(n_s)}(z'', f_i) = -\alpha^{(n_s)}(z'', f_i) - j\beta^{(n_s)}(z'', f_i) \quad (13)$$

The SPM and XPM islands are located on the two axes with the feature that,

$$\{f_{s,m_{ch}} \leq f_1 \leq f_{e,m_{ch}}, f_{s,\text{CUT}} \leq f_2 \leq f_{e,\text{CUT}}, f_{s,m_{ch}} \leq f_3 \leq f_{e,m_{ch}}\}$$
$$\text{or } \{f_{s,\text{CUT}} \leq f_1 \leq f_{e,\text{CUT}}, f_{s,m_{ch}} \leq f_2 \leq f_{e,m_{ch}}, f_{s,m_{ch}} \leq f_3 \leq f_{e,m_{ch}}\} \quad (14)$$

Therefore, the term $\kappa^{(n_s)}(z'',f_1) + \kappa^{(n_s)}(z'',f_2) + \kappa^{*(n_s)}(z'',f_3) - \kappa^{(n_s)}(z'',f_1+f_2-f_3)$ can be simplified as,

$$\kappa^{(n_s)}(z'', f_1) + \kappa^{(n_s)}(z'', f_2) + \kappa^{*(n_s)}(z'', f_3) - \kappa^{(n_s)}(z'', f_1+f_2-f_3)$$
$$= -2\alpha^{(n_s)}(z'', f_{c,m_{ch}}) + j \cdot \chi^{(n_s)}(f_1, f_2, f) \quad (15)$$

Where we assume that the loss within each channel is constant and identical, equating to the value at the center frequency. The dispersion is extended to the 4-th order to better model the actual performance in UWB systems as,

$$\beta^{(n_s)}(z'', f_i) \approx \beta_0^{(n_s)} + 2\pi\beta_1^{(n_s)}\left(f_i - f_{\text{ref}}^{(n_s)}\right) + 2\pi^2 \beta_2^{(n_s)}\left(f_i - f_{\text{ref}}^{(n_s)}\right)^2$$
$$+ \frac{4\pi^3}{3}\beta_3^{(n_s)}\left(f_i - f_{\text{ref}}^{(n_s)}\right)^3 + \frac{2}{3}\pi^4 \beta_4^{(n_s)}\left(f_i - f_{\text{ref}}^{(n_s)}\right)^4 \quad (16)$$

Where $f_{\text{ref}}^{(n_s)}$ is the frequency at which $\beta_i^{(n_s)}, i = 1, 2, 3, 4$ are measured. Substituting Eq. (16) back to Eq. (15) gives,

$$\chi^{(n_s)}(f_1, f_2, f) = 4\pi^2 (f_1 - f)(f_2 - f) \cdot \left\{ \beta_2^{(n_s)} + \pi \beta_3^{(n_s)} \left( f_1 + f_2 - 2 f_{\text{ref}}^{(n_s)} \right) \right.$$
$$\left. + \frac{2}{3} \pi^2 \beta_4^{(n_s)} \left[ \left( f_1 - f_{\text{ref}}^{(n_s)} \right)^2 + \left( f_1 - f_{\text{ref}}^{(n_s)} \right)\left( f_2 - f_{\text{ref}}^{(n_s)} \right) - \left( f_2 - f_{\text{ref}}^{(n_s)} \right)^2 \right] \right\} \quad (17)$$

Up to now, the IGN model can be expressed as,

$$G_{\text{NLI}}(f) \approx \frac{16}{27} \sum_{n_s=1}^{N_s} \left[ \gamma^{(n_s)}\left( f_{\text{CUT}}, f_{m_{\text{ch}}} \right) \right]^2 \sum_{m_{\text{ch}}=1}^{N_{\text{ch}}} G_{\text{CUT}}^{(n_{s,\text{st}})} \left( G_{m_{\text{ch}}}^{(n_{s,\text{st}})} \right)^2 \left( 2 - \delta_{\text{CUT}, m_{\text{ch}}} \right) \Gamma_{n_{s,\text{st}}}^{N_s}(f_{\text{CUT}})$$
$$\iint_{S(m_{\text{ch}}, m_{\text{ch}}, \text{CUT})} \left| \int_0^{L_s^{(n_s)}} e^{\int_0^{z'} \left[ -2\alpha^{(n_s)}\left( z'', f_{c, m_{\text{ch}}} \right) + j \cdot \chi^{(n_s)}(f_1, f_2, f) \right] dz''} dz' \right|^2 df_1 df_2 \quad (18)$$

Where $G_{\text{CUT}/m_{\text{ch}}}^{(n_{s,\text{st}})}$ is the signal PSD at the start of the $n_s$-th span, obtained by propagating the signal PSD at the start of the link to the start of the $n_s$-th span. $\Gamma_{n_{s,\text{st}}}^{N_s}(f_{\text{CUT}})$ is the accumulated loss/gain on CUT from the start of $n_s$-th span to the end of the link, signifying the linear accumulation of the NLI after its generation in the $n_s$-th span. If all spans are transparent such that the span loss is completely compensated by gain, the IGN model can be simplified as,

$$G_{\text{NLI}}(f) \approx \frac{16}{27} \sum_{n_s=1}^{N_s} \left[ \gamma^{(n_s)}\left( f_{\text{CUT}}, f_{m_{\text{ch}}} \right) \right]^2 \sum_{m_{\text{ch}}=1}^{N_{\text{ch}}} G_{\text{CUT}} \left( G_{m_{\text{ch}}} \right)^2 \left( 2 - \delta_{\text{CUT}, m_{\text{ch}}} \right)$$
$$\iint_{S(m_{\text{ch}}, m_{\text{ch}}, \text{CUT})} \left| \int_0^{L_s^{(n_s)}} e^{\int_0^{z'} \left[ -2\alpha^{(n_s)}\left( z'', f_{c, m_{\text{ch}}} \right) + j \cdot \chi^{(n_s)}(f_1, f_2, f) \right] dz''} dz' \right|^2 df_1 df_2 \quad (19)$$

In realistic scenarios, each span may encounter varying levels of gain/loss. Hence, we use Eq. (18) to derive CFM6. In sect. 2, we divided each span into two segments to model the fiber field loss. Accordingly, we separately find out the CFMs for them by assuming that they are completely independent from each other. Eq. (18) can then be split into two distinct contributions,

$$G_{\text{NLI}}(f) \approx \frac{16}{27} \sum_{n_s=1}^{N_s} \left[ \gamma^{(n_s)}\left( f_{\text{CUT}}, f_{m_{\text{ch}}} \right) \right]^2 \sum_{m_{\text{ch}}=1}^{N_{\text{ch}}} G_{\text{CUT}}^{(n_{s,\text{st}})} \left( G_{m_{\text{ch}}}^{(n_{s,\text{st}})} \right)^2 \left( 2 - \delta_{\text{CUT}, m_{\text{ch}}} \right) \Gamma_{n_{s,\text{st}}}^{N_s}(f_{\text{CUT}})$$
$$\iint_{S(m_{\text{ch}}, m_{\text{ch}}, \text{CUT})} \left| \int_0^{L_{s,\text{st}}^{(n_s)}} e^{\int_0^{z'} \left[ -2\alpha^{(n_s)}\left( z'', f_{c, m_{\text{ch}}} \right) + j \cdot \chi^{(n_s)}(f_1, f_2, f) \right] dz''} dz' \right|^2 df_1 df_2$$
$$+ \frac{16}{27} \sum_{n_s=1}^{N_s} \left[ \gamma^{(n_s)}\left( f_{\text{CUT}}, f_{m_{\text{ch}}} \right) \right]^2 \sum_{m_{\text{ch}}=1}^{N_{\text{ch}}} G_{\text{CUT}}^{(n_{s,\text{st}})} \left( G_{m_{\text{ch}}}^{(n_{s,\text{st}})} \right)^2 \left( 2 - \delta_{\text{CUT}, m_{\text{ch}}} \right) \Gamma_{n_{s,\text{st}}}^{N_s}(f_{\text{CUT}}) \quad (20)$$
$$\iint_{S(m_{\text{ch}}, m_{\text{ch}}, \text{CUT})} \left| \int_{L_{s,\text{st}}^{(n_s)}}^{L_s^{(n_s)}} e^{\int_0^{z'} \left[ -2\alpha^{(n_s)}\left( z'', f_{c, m_{\text{ch}}} \right) + j \cdot \chi^{(n_s)}(f_1, f_2, f) \right] dz''} dz' \right|^2 df_1 df_2$$
$$\triangleq G_{\text{NLI,st}}(f) + G_{\text{NLI,end}}(f)$$

### 3.1 the first contribution in CFM6

The first contribution to NLI PSD from the initial segment is expressed as,

$$G_{\text{NLI,st}}(f) \approx \frac{16}{27} \sum_{n_s=1}^{N_s} \left[\gamma^{(n_s)}\left(f_{\text{CUT}}, f_{m_{\text{ch}}}\right)\right]^2 \sum_{m_{\text{ch}}=1}^{N_{\text{ch}}} G_{\text{CUT}}^{(n_{s,\text{st}})} \left(G_{m_{\text{ch}}}^{(n_{s,\text{st}})}\right)^2 \left(2-\delta_{\text{CUT},m_{\text{ch}}}\right) \Gamma_{n_{s,\text{st}}}^{N_s}(f_{\text{CUT}})$$

$$\iint_{S(m_{\text{ch}},m_{\text{ch}},\text{CUT})} \left| \int_0^{L_{s,\text{st}}^{(n_s)}} e^{\int_0^{z'} \left[-2\alpha_{\text{st}}^{(n_s)}(z'',f_{c,m_{\text{ch}}})+j\cdot\chi^{(n_s)}(f_1,f_2,f)\right]dz''} dz' \right|^2 df_1 df_2 \quad (21)$$

Taking Eq. (6) into Eq. (21) gives,

$$G_{\text{NLI,st}}(f) \approx \frac{16}{27} \sum_{n_s=1}^{N_s} \left[\gamma^{(n_s)}\left(f_{\text{CUT}}, f_{m_{\text{ch}}}\right)\right]^2 \sum_{m_{\text{ch}}=1}^{N_{\text{ch}}} G_{\text{CUT}}^{(n_{s,\text{st}})} \left(G_{m_{\text{ch}}}^{(n_{s,\text{st}})}\right)^2 \left(2-\delta_{\text{CUT},m_{\text{ch}}}\right) \Gamma_{n_{s,\text{st}}}^{N_s}(f_{\text{CUT}}) \cdot e^{\frac{-4\alpha_{1,\text{st}}^{(n_s)}(f_{c,m_{\text{ch}}})}{\sigma_{\text{st}}^{(n_s)}(f_{c,m_{\text{ch}}})}}$$

$$\iint_{S(m_{\text{ch}},m_{\text{ch}},\text{CUT})} \left| \int_0^{L_{s,\text{st}}^{(n_s)}} e^{\frac{2\alpha_{1,\text{st}}^{(n_s)}(f_{c,m_{\text{ch}}})}{\sigma_{\text{st}}^{(n_s)}(f_{c,m_{\text{ch}}})} \exp\left(-\sigma_{\text{st}}^{(n_s)}(f_{c,m_{\text{ch}}})z'\right)} e^{-2\alpha_{0,\text{st}}^{(n_s)}(f_{c,m_{\text{ch}}})z'+j\cdot\chi^{(n_s)}(f_1,f_2,f)z'} dz' \right|^2 df_1 df_2 \quad (22)$$

We express the first term within the single integral on $z'$ by a series expansion,

$$e^{\frac{2\alpha_{1,\text{st}}^{(n_s)}(f_{c,m_{\text{ch}}})}{\sigma_{\text{st}}^{(n_s)}(f_{c,m_{\text{ch}}})} \exp\left(-\sigma_{\text{st}}^{(n_s)}(f_{c,m_{\text{ch}}})z'\right)} = \sum_{n=0}^{\infty} \frac{1}{n!} \left[\frac{2\alpha_{1,\text{st}}^{(n_s)}(f_{c,m_{\text{ch}}})}{\sigma_{\text{st}}^{(n_s)}(f_{c,m_{\text{ch}}})} \exp\left(-\sigma_{\text{st}}^{(n_s)}(f_{c,m_{\text{ch}}})z'\right)\right]^n$$

$$\approx \sum_{n=0}^{M_{\text{st}}^{(n_s)}(f_{c,m_{\text{ch}}})} \frac{1}{n!} \left[\frac{2\alpha_{1,\text{st}}^{(n_s)}(f_{c,m_{\text{ch}}})}{\sigma_{\text{st}}^{(n_s)}(f_{c,m_{\text{ch}}})}\right]^n e^{-n\sigma_{\text{st}}^{(n_s)}(f_{c,m_{\text{ch}}})z'}, M_{\text{st}}^{(n_s)}(f_{c,m_{\text{ch}}}) = \left\lfloor 10 \cdot \left|\frac{2\alpha_{1,\text{st}}^{(n_s)}(f_{c,m_{\text{ch}}})}{\sigma_{\text{st}}^{(n_s)}(f_{c,m_{\text{ch}}})}\right| \right\rfloor \quad (23)$$

Where we approximate the infinite terms by considering only the first $M_{\text{st}}^{(n_s)}(f_{c,m_{\text{ch}}})+1$ terms. The $M_{\text{st}}^{(n_s)}(f_{c,m_{\text{ch}}})+1$-th term is significantly larger than the subsequent terms to make them negligible. Substituting it back to Eq. (22) gives,

$$G_{\text{NLI,st}}(f) \approx \frac{16}{27} \sum_{n_s=1}^{N_s} \left[\gamma^{(n_s)}\left(f_{\text{CUT}}, f_{m_{\text{ch}}}\right)\right]^2 \sum_{m_{\text{ch}}=1}^{N_{\text{ch}}} G_{\text{CUT}}^{(n_{s,\text{st}})} \left(G_{m_{\text{ch}}}^{(n_{s,\text{st}})}\right)^2 \left(2-\delta_{\text{CUT},m_{\text{ch}}}\right) \Gamma_{n_{s,\text{st}}}^{N_s}(f_{\text{CUT}}) e^{\frac{-4\alpha_{1,\text{st}}^{(n_s)}(f_{c,m_{\text{ch}}})}{\sigma_{\text{st}}^{(n_s)}(f_{c,m_{\text{ch}}})}}$$

$$\iint_{S(m_{\text{ch}},m_{\text{ch}},\text{CUT})} \left| \sum_{n=0}^{M_{\text{st}}^{(n_s)}(f_{c,m_{\text{ch}}})} \frac{1}{n!} \left[\frac{2\alpha_{1,\text{st}}^{(n_s)}(f_{c,m_{\text{ch}}})}{\sigma_{\text{st}}^{(n_s)}(f_{c,m_{\text{ch}}})}\right]^n \frac{e^{L_{s,\text{st}}^{(n_s)}\left[-2\alpha_{0,\text{st}}^{(n_s)}(f_{c,m_{\text{ch}}})-n\sigma_{\text{st}}^{(n_s)}(f_{c,m_{\text{ch}}})+j\cdot\chi^{(n_s)}(f_1,f_2,f)\right]}-1}{-2\alpha_{0,\text{st}}^{(n_s)}(f_{c,m_{\text{ch}}})-n\sigma_{\text{st}}^{(n_s)}(f_{c,m_{\text{ch}}})+j\cdot\chi^{(n_s)}(f_1,f_2,f)} \right|^2 df_1 df_2 \quad (24)$$

Generally, we could assume $e^{L_{s,\text{st}}^{(n_s)}\left[-2\alpha_{0,\text{st}}^{(n_s)}(f_{c,m_{\text{ch}}})-n\sigma_{\text{st}}^{(n_s)}(f_{c,m_{\text{ch}}})+j\cdot\chi^{(n_s)}(f_1,f_2,f)\right]} \ll 1$ to get,

$$G_{\text{NLI,st}}(f) \approx \frac{16}{27} \sum_{n_s=1}^{N_s} \left[\gamma^{(n_s)}\left(f_{\text{CUT}}, f_{m_{\text{ch}}}\right)\right]^2 \sum_{m_{\text{ch}}=1}^{N_{\text{ch}}} G_{\text{CUT}}^{(n_{s,\text{st}})} \left(G_{m_{\text{ch}}}^{(n_{s,\text{st}})}\right)^2 \left(2-\delta_{\text{CUT},m_{\text{ch}}}\right) \Gamma_{n_{s,\text{st}}}^{N_s}(f_{\text{CUT}})$$

$$\cdot e^{\frac{-4\alpha_{1,\text{st}}^{(n_s)}(f_{c,m_{\text{ch}}})}{\sigma_{\text{st}}^{(n_s)}(f_{c,m_{\text{ch}}})}} \sum_{n_1=0}^{M_{\text{st}}^{(n_s)}(f_{c,m_{\text{ch}}})} \sum_{n_2=0}^{M_{\text{st}}^{(n_s)}(f_{c,m_{\text{ch}}})} \frac{2}{n_1! n_2!} \left[\frac{2\alpha_{1,\text{st}}^{(n_s)}(f_{c,m_{\text{ch}}})}{\sigma_{\text{st}}^{(n_s)}(f_{c,m_{\text{ch}}})}\right]^{n_1+n_2} \cdot \frac{1}{4\alpha_{0,\text{st}}^{(n_s)}(f_{c,m_{\text{ch}}})+(n_1+n_2)\sigma_{\text{st}}^{(n_s)}(f_{c,m_{\text{ch}}})} \quad (25)$$

$$\iint_{S(m_{\text{ch}},m_{\text{ch}},\text{CUT})} \frac{2\alpha_{0,\text{st}}^{(n_s)}(f_{c,m_{\text{ch}}})+n_1\sigma_{\text{st}}^{(n_s)}(f_{c,m_{\text{ch}}})}{\left[2\alpha_{0,\text{st}}^{(n_s)}(f_{c,m_{\text{ch}}})+n_1\sigma_{\text{st}}^{(n_s)}(f_{c,m_{\text{ch}}})\right]^2+\chi^2} df_1 df_2$$

To fully integrate Eq. (25), we further make some assumptions and calculations as in CFM5 (Eq. (79) to (85) in [11]), and then obtain the first contribution in CFM6,

$$G_{\text{NLI,st}}(f_{\text{CUT}}) \approx \frac{16}{27} \sum_{n_s=1}^{N_s} \left[\gamma^{(n_s)}(f_{\text{CUT}}, f_{m_{\text{ch}}})\right]^2 \sum_{m_{\text{ch}}=1}^{N_{\text{ch}}} G_{\text{CUT}}^{(n_{s,\text{st}})} \left(G_{m_{\text{ch}}}^{(n_{s,\text{st}})}\right)^2 (2-\delta_{\text{CUT},m_{\text{ch}}}) \Gamma_{n_{s,\text{st}}}^{N_s}(f_{\text{CUT}})$$

$$\cdot e^{\frac{-4\alpha_{1,\text{st}}^{(n_s)}(f_{c,m_{\text{ch}}})}{\sigma_{\text{st}}^{(n_s)}(f_{c,m_{\text{ch}}})}} \cdot \frac{1}{4\pi\beta_{2,\text{eff}}^{(n_s)}(f_{c,m_{\text{ch}}}, f_{c,\text{CUT}})} \sum_{n_1=0}^{M_{\text{st}}^{(n_s)}(f_{c,m_{\text{ch}}})} \sum_{n_2=0}^{M_{\text{st}}^{(n_s)}(f_{c,m_{\text{ch}}})} \frac{2}{n_1! n_2!} \left[\frac{2\alpha_{1,\text{st}}^{(n_s)}(f_{c,m_{\text{ch}}})}{\sigma_{\text{st}}^{(n_s)}(f_{c,m_{\text{ch}}})}\right]^{n_1+n_2}$$

$$\cdot \frac{1}{4\alpha_{0,\text{st}}^{(n_s)}(f_{c,m_{\text{ch}}}) + (n_1+n_2)\sigma_{\text{st}}^{(n_s)}(f_{c,m_{\text{ch}}})}$$

$$\cdot \left\{ \operatorname{asinh}\left(\frac{\pi^2 \beta_{2,\text{eff}}^{(n_s)}(f_{c,m_{\text{ch}}}, f_{c,\text{CUT}}) \cdot \text{BW}_{\text{CUT}} \cdot (f_{e,m_{\text{ch}}} - f_{c,\text{CUT}})}{2\alpha_{0,\text{st}}^{(n_s)}(f_{c,m_{\text{ch}}}) + n_1 \sigma_{\text{st}}^{(n_s)}(f_{c,m_{\text{ch}}})}\right) \right.$$

$$\left. -\operatorname{asinh}\left(\frac{\pi^2 \beta_{2,\text{eff}}^{(n_s)}(f_{c,m_{\text{ch}}}, f_{c,\text{CUT}}) \cdot \text{BW}_{\text{CUT}} \cdot (f_{s,m_{\text{ch}}} - f_{c,\text{CUT}})}{2\alpha_{0,\text{st}}^{(n_s)}(f_{c,m_{\text{ch}}}) + n_1 \sigma_{\text{st}}^{(n_s)}(f_{c,m_{\text{ch}}})}\right) \right\} \quad (26)$$

### 3.2 the second contribution in CFM6

The second contribution to NLI PSD from the second segment is expressed as,

$$G_{\text{NLI,end}}(f) \approx \frac{16}{27} \sum_{n_s=1}^{N_s} \left[\gamma^{(n_s)}(f_{\text{CUT}}, f_{m_{\text{ch}}})\right]^2 \sum_{m_{\text{ch}}=1}^{N_{\text{ch}}} G_{\text{CUT}}^{(n_{s,\text{st}})} \left(G_{m_{\text{ch}}}^{(n_{s,\text{st}})}\right)^2 (2-\delta_{\text{CUT},m_{\text{ch}}}) \Gamma_{n_{s,\text{st}}}^{N_s}(f_{\text{CUT}})$$

$$\iint_{S(m_{\text{ch}}, m_{\text{ch}}, \text{CUT})} \left| \int_{L_{s,\text{st}}^{(n_s)}}^{L_s^{(n_s)}} e^{\int_0^{z'} \left[-2\alpha^{(n_s)}(z'', f_{c,m_{\text{ch}}}) + j\cdot\chi^{(n_s)}(f_1, f_2, f)\right] dz''} dz' \right|^2 df_1 df_2 \quad (27)$$

Initially, our attention is directed towards the integrand within the double integral.

$$\left| \int_{L_{s,\text{st}}^{(n_s)}}^{L_s^{(n_s)}} e^{\int_0^{z'}\left[-2\alpha^{(n_s)}(z'', f_{c,m_{\text{ch}}}) + j\cdot\chi^{(n_s)}(f_1, f_2, f)\right] dz''} dz' \right|^2$$

$$= \left| \int_{L_{s,\text{st}}^{(n_s)}}^{L_s^{(n_s)}} e^{\int_0^{L_{s,\text{st}}^{(n_s)}}\left[-2\alpha_{\text{st}}^{(n_s)}(z'', f_{c,m_{\text{ch}}})\right] dz''} e^{\int_{L_{s,\text{st}}^{(n_s)}}^{L_s^{(n_s)}}\left[-2\alpha_{\text{end}}^{(n_s)}(z'', f_{c,m_{\text{ch}}})\right] dz''} e^{j\cdot\chi^{(n_s)}(f_1, f_2, f) z'} dz' \right|^2$$

$$= \left| \frac{p^{(n_s)}(L_{s,\text{st}}^{(n_s)}, f_{c,m_{\text{ch}}})}{p^{(n_s)}(0, f_{c,m_{\text{ch}}})} \cdot \int_0^{L_{s,\text{end}}^{(n_s)}} \frac{p^{(n_s)}(z' + L_{s,\text{st}}^{(n_s)}, f_{c,m_{\text{ch}}})}{p^{(n_s)}(L_{s,\text{st}}^{(n_s)}, f_{c,m_{\text{ch}}})} \cdot e^{j\cdot\chi^{(n_s)}(f_1, f_2, f)(z' + L_{s,\text{st}}^{(n_s)})} dz' \right|^2 \quad (28)$$

$$= \left| \frac{1}{p^{(n_s)}(0, f_{c,m_{\text{ch}}})} \cdot \int_0^{L_{s,\text{end}}^{(n_s)}} p^{(n_s)}(z' + L_{s,\text{st}}^{(n_s)}, f_{c,m_{\text{ch}}}) \cdot e^{j\cdot\chi^{(n_s)}(f_1, f_2, f)(z' + L_{s,\text{st}}^{(n_s)})} dz' \right|^2$$

Above, we move the starting point to z=0. Next, we substitute the variable $z'$ with $t' = L_{s,\text{end}}^{(n_s)} - z'$ to facilitate the reversal for employing the loss model in Eq. (7).

$$\left| \int_{L_{s,st}^{(n_s)}}^{L_s^{(n_s)}} e^{\int_0^{z'} \left[ -2\alpha^{(n_s)}\left(z'', f_{c,m_{ch}}\right) + j\cdot\chi^{(n_s)}(f_1,f_2,f) \right]dz''} dz' \right|^2$$

$$= \left| -\frac{1}{p^{(n_s)}\left(0, f_{c,m_{ch}}\right)} \int_{L_{s,end}^{(n_s)}}^{0} p^{(n_s)}\left(L_s^{(n_s)} - t', f_{c,m_{ch}}\right) \cdot e^{j\cdot\chi^{(n_s)}(f_1,f_2,f)\left(L_s^{(n_s)} - t'\right)} dt' \right|^2 \quad (29)$$

$$= \left| \frac{1}{p^{(n_s)}\left(0, f_{c,m_{ch}}\right)} \int_0^{L_{s,end}^{(n_s)}} P_{\text{end-flip}}^{(n_s)}\left(z', f_{c,m_{ch}}\right) \cdot e^{j\cdot\chi^{(n_s)}(f_1,f_2,f)\left(L_s^{(n_s)} - z'\right)} dz' \right|^2$$

Now, we can substitute Eq. (8) into Eq. (29), and subsequently substitute the result into Eq. (27) to obtain,

$$G_{\text{NLI,end}}(f) \approx \frac{16}{27} \sum_{n_s=1}^{N_s} \left[ \gamma^{(n_s)}\left(f_{\text{CUT}}, f_{m_{ch}}\right) \right]^2 \sum_{m_{ch}=1}^{N_{ch}} G_{\text{CUT}}^{(n_{s,st})} \left( G_{m_{ch}}^{(n_{s,st})} \right)^2 \left(2 - \delta_{\text{CUT}, m_{ch}}\right) \Gamma_{n_{s,st}}^{N_s}(f_{\text{CUT}}) e^{\frac{-4\alpha_{1,\text{end,flip}}^{(n_s)}\left(f_{c,m_{ch}}\right)}{\sigma_{\text{end,flip}}^{(n_s)}\left(f_{c,m_{ch}}\right)}}$$

$$\iint\limits_{S(m_{ch},m_{ch},\text{CUT})} \left| \frac{p^{(n_s)}\left(L_s, f_{c,m_{ch}}\right)}{p^{(n_s)}\left(0, f_{c,m_{ch}}\right)} \int_0^{L_{s,end}^{(n_s)}} e^{\frac{2\alpha_{1,\text{end,flip}}^{(n_s)}\left(f_{c,m_{ch}}\right)}{\sigma_{\text{end,flip}}^{(n_s)}\left(f_{c,m_{ch}}\right)} \exp\left(-\sigma_{\text{end,flip}}^{(n_s)}\left(f_{c,m_{ch}}\right)\cdot z'\right)} \cdot e^{-2\alpha_{0,\text{end,flip}}^{(n_s)}\left(f_{c,m_{ch}}\right) z' - j\cdot\chi^{(n_s)}(f_1,f_2,f) z'} dz' \right|^2 df_1 df_2$$

$$= \frac{16}{27} \sum_{n_s=1}^{N_s} \left(\gamma^{(n_s)}\right)^2 \sum_{m_{ch}=1}^{N_{ch}} G_{\text{CUT}}^{(n_{s,end})} \left(G_{m_{ch}}^{(n_{s,end})}\right)^2 \left(2 - \delta_{\text{CUT}, m_{ch}}\right) \Gamma_{n_{s,end}}^{N_s}(f_{\text{CUT}}) e^{\frac{-4\alpha_{1,\text{end}}^{(n_s)}\left(f_{c,m_{ch}}\right)}{\sigma_{\text{end}}^{(n_s)}\left(f_{c,m_{ch}}\right)}}$$

$$\iint\limits_{S(m_{ch},m_{ch},\text{CUT})} \left| \int_0^{L_{s,end}^{(n_s)}} e^{\frac{2\alpha_{1,\text{end}}^{(n_s)}\left(f_{c,m_{ch}}\right)}{\sigma_{\text{end}}^{(n_s)}\left(f_{c,m_{ch}}\right)} \exp\left(-\sigma_{\text{end}}^{(n_s)}\left(f_{c,m_{ch}}\right)\cdot z'\right)} \cdot e^{-2\alpha_{0,\text{end}}^{(n_s)}\left(f_{c,m_{ch}}\right) z' - j\cdot\chi^{(n_s)}(f_1,f_2,f) z'} dz' \right|^2 df_1 df_2$$

(30)

For simplicity, we have already removed the 'flip' in the subscripts of the three parameters $\alpha_{0,\text{end,flip}}^{(n_s)}, \alpha_{1,\text{end,flip}}^{(n_s)}, \sigma_{\text{end}}^{(n_s)}$ in Eq. (30). Then, we expand the first term using a Taylor series,

$$e^{\frac{2\alpha_{1,\text{end}}^{(n_s)}\left(f_{c,m_{ch}}\right)}{\sigma_{\text{end}}^{(n_s)}\left(f_{c,m_{ch}}\right)} \exp\left(-\sigma_{\text{end}}^{(n_s)}\left(f_{c,m_{ch}}\right)\cdot z'\right)} = \sum_{n=0}^{\infty} \frac{1}{n!} \left[ \frac{2\alpha_{1,\text{end}}^{(n_s)}\left(f_{c,m_{ch}}\right)}{\sigma_{\text{end}}^{(n_s)}\left(f_{c,m_{ch}}\right)} \exp\left(-\sigma_{\text{end}}^{(n_s)}\left(f_{c,m_{ch}}\right) z'\right) \right]^n$$

$$\approx \sum_{n=0}^{M_{\text{end}}^{(n_s)}\left(f_{c,m_{ch}}\right)} \frac{1}{n!} \left[ \frac{2\alpha_{1,\text{end}}^{(n_s)}\left(f_{c,m_{ch}}\right)}{\sigma_{\text{end}}^{(n_s)}\left(f_{c,m_{ch}}\right)} \right]^n e^{-n\sigma_{\text{end}}^{(n_s)}\left(f_{c,m_{ch}}\right) z'}, M_{\text{end}}^{(n_s)}\left(f_{c,m_{ch}}\right) = \left\lfloor 10 \cdot \left| \frac{2\alpha_{1,\text{end}}^{(n_s)}\left(f_{c,m_{ch}}\right)}{\sigma_{\text{end}}^{(n_s)}\left(f_{c,m_{ch}}\right)} \right| \right\rfloor$$

(31)

It repeats the procedure outlined in Eq. (23), replacing the infinite terms with the first $M_{\text{end}}^{(n_s)}\left(f_{c,m_{ch}}\right) + 1$ terms. Substituting it back to Eq. (30) gives,

$$G_{\text{NLI,end}}(f) \approx \frac{16}{27} \sum_{n_s=1}^{N_s} \left[\gamma^{(n_s)}\left(f_{\text{CUT}}, f_{m_{\text{ch}}}\right)\right]^2 \sum_{m_{\text{ch}}=1}^{N_{\text{ch}}} G_{\text{CUT}}^{(n_{s,\text{end}})} \left(G_{m_{\text{ch}}}^{(n_{s,\text{end}})}\right)^2 \left(2-\delta_{\text{CUT},m_{\text{ch}}}\right) \Gamma_{n_{s,\text{end}}}^{N_s}(f_{\text{CUT}}) e^{\frac{-4\alpha_{1,\text{end}}^{(n_s)}(f_{c,m_{\text{ch}}})}{\sigma_{\text{end}}^{(n_s)}(f_{c,m_{\text{ch}}})}}$$

$$\iint_{S(m_{\text{ch}},m_{\text{ch}},\text{CUT})} \left| \sum_{n=0}^{M_{\text{end}}^{(n_s)}(f_{c,m_{\text{ch}}})} \frac{1}{n!} \left[\frac{2\alpha_{1,\text{end}}^{(n_s)}(f_{c,m_{\text{ch}}})}{\sigma_{\text{end}}^{(n_s)}(f_{c,m_{\text{ch}}})}\right]^n \frac{e^{L_{s,\text{end}}^{(n_s)}\left[-2\alpha_{0,\text{end}}^{(n_s)}(f_{c,m_{\text{ch}}})-n\sigma_{\text{end}}^{(n_s)}(f_{c,m_{\text{ch}}})+j\cdot\chi^{(n_s)}(f_1,f_2,f)\right]}-1}{-2\alpha_{0,\text{end}}^{(n_s)}(f_{c,m_{\text{ch}}})-n\sigma_{\text{end}}^{(n_s)}(f_{c,m_{\text{ch}}})+j\cdot\chi^{(n_s)}(f_1,f_2,f)} \right|^2 df_1 df_2$$

(32)

Again, we assume that $e^{L_{s,\text{end}}^{(n_s)}\left[-2\alpha_{0,\text{end}}^{(n_s)}(f_{c,m_{\text{ch}}})-n\sigma_{\text{end}}^{(n_s)}(f_{c,m_{\text{ch}}})+j\cdot\chi^{(n_s)}(f_1,f_2,f)\right]} \ll 1$. However, this assumption may not hold universally, particularly when $n=0$ as the parameter $\alpha_{0,\text{end}}^{(n_s)}(f_{c,m_{\text{ch}}})$ can become negative for many channels. To solve the problem stemming from the zero crossing, we constrain the range of $\sigma_{\text{end}}^{(n_s)}(f_{c,m_{\text{ch}}})$ such that all $\alpha_{0,\text{end}}^{(n_s)}(f_{c,m_{\text{ch}}})$ are negative with magnitudes close to zero.

To fully integrate Eq. (32), we further make some assumptions and calculations as in CFM5 (Eq. (79) to (85) in [11]), and then obtain the second contributions in CFM6,

$$G_{\text{NLI,end}}(f) \approx \frac{16}{27} \sum_{n_s=1}^{N_s} \left[\gamma^{(n_s)}\left(f_{\text{CUT}}, f_{m_{\text{ch}}}\right)\right]^2 \sum_{m_{\text{ch}}=1}^{N_{\text{ch}}} G_{\text{CUT}}^{(n_{s,\text{end}})} \left(G_{m_{\text{ch}}}^{(n_{s,\text{end}})}\right)^2 \left(2-\delta_{\text{CUT},m_{\text{ch}}}\right) \Gamma_{n_{s,\text{end}}}^{N_s}(f_{\text{CUT}})$$

$$\cdot e^{\frac{-4\alpha_{1,\text{end}}^{(n_s)}(f_{c,m_{\text{ch}}})}{\sigma_{\text{end}}^{(n_s)}(f_{c,m_{\text{ch}}})}} \frac{1}{4\pi\beta_{2,\text{eff}}^{(n_s)}(f_{c,m_{\text{ch}}}, f_{c,\text{CUT}})} \sum_{n_1=0}^{M_{\text{end}}^{(n_s)}(f_{c,m_{\text{ch}}})} \sum_{n_2=0}^{M_{\text{end}}^{(n_s)}(f_{c,m_{\text{ch}}})} \frac{2}{n_1!n_2!} \left[\frac{2\alpha_{1,\text{end}}^{(n_s)}(f_{c,m_{\text{ch}}})}{\sigma_{\text{end}}^{(n_s)}(f_{c,m_{\text{ch}}})}\right]^{n_1+n_2}$$

$$\cdot \frac{1}{4\alpha_{0,\text{end}}^{(n_s)}(f_{c,m_{\text{ch}}}) + (n_1+n_2)\sigma_{\text{end}}^{(n_s)}(f_{c,m_{\text{ch}}})}$$

$$\cdot \left\{ \text{asinh}\left(\frac{\pi^2 \beta_{2,\text{eff}}^{(n_s)}(f_{c,m_{\text{ch}}}, f_{c,\text{CUT}}) \cdot \text{BW}_{\text{CUT}} \cdot (f_{e,m_{\text{ch}}} - f_{c,\text{CUT}})}{2\alpha_{0,\text{end}}^{(n_s)}(f_{c,m_{\text{ch}}}) + n_1\sigma_{\text{end}}^{(n_s)}(f_{c,m_{\text{ch}}})}\right) \right.$$

$$\left. -\text{asinh}\left(\frac{\pi^2 \beta_{2,\text{eff}}^{(n_s)}(f_{c,m_{\text{ch}}}, f_{c,\text{CUT}}) \cdot \text{BW}_{\text{CUT}} \cdot (f_{s,m_{\text{ch}}} - f_{c,\text{CUT}})}{2\alpha_{0,\text{end}}^{(n_s)}(f_{c,m_{\text{ch}}}) + n_1\sigma_{\text{end}}^{(n_s)}(f_{c,m_{\text{ch}}})}\right) \right\}$$

(33)

Which is similar to Eq. (26). If the second segment is characterized from the backward direction, with $z = 0$ at the span end, for both the signal and NLI generation, that is,

- The signal PSD is $G_{\text{CUT}}^{(n_{s,\text{end}})}, G_{m_{\text{ch}}}^{(n_{s,\text{end}})}$ for the channel CUT, $m_{\text{ch}}$ launched into the $n_s$-th span.

- The accumulated gain for NLI starts from the end of the span to the end of the link, denoted as $\Gamma_{n_{s,\text{end}}}^{N_s}(f_{\text{CUT}})$.

- The three parameters in the loss model are exactly $\alpha_{0,\text{end}}^{(n_s)}(f_{c,m_{\text{ch}}}), \alpha_{1,\text{end}}^{(n_s)}(f_{c,m_{\text{ch}}}), \sigma_{\text{end}}^{(n_s)}(f_{c,m_{\text{ch}}})$.

This contribution in Eq. (33) can be directly found through the same procedure as for the first contribution.

### 3.3 IGN to EGN

CFM5 additionally incorporates a machine-learning-based 'correction term' to enhance the approximation of the EGN model from IGN. All the machine-learning factors are shown in Table IV in [15]. In CFM6, we take the same parameter $\rho_{\text{ML}}^{(n_s)}$ and obtain,

- The first contribution:

$$G_{\text{NLI,st}}(f_{\text{CUT}}) \approx \frac{16}{27} \sum_{n_s=1}^{N_s} \left[\gamma^{(n_s)}(f_{\text{CUT}}, f_{m_{\text{ch}}})\right]^2 \sum_{m_{\text{ch}}=1}^{N_{\text{ch}}} G_{\text{CUT}}^{(n_{s,\text{st}})} \left(G_{m_{\text{ch}}}^{(n_{s,\text{st}})}\right)^2 (2-\delta_{\text{CUT},m_{\text{ch}}})\left(\rho_{\text{ML}}^{(n_s)}\right)\Gamma_{n_{s,\text{st}}}^{N_s}(f_{\text{CUT}})$$

$$\cdot e^{\frac{-4\alpha_{1,\text{st}}^{(n_s)}(f_{c,m_{\text{ch}}})}{\sigma_{\text{st}}^{(n_s)}(f_{c,m_{\text{ch}}})}} \cdot \frac{1}{4\pi\beta_{2,\text{eff}}^{(n_s)}(f_{c,m_{\text{ch}}}, f_{c,\text{CUT}})} \sum_{n_1=0}^{M_{\text{st}}^{(n_s)}(f_{c,m_{\text{ch}}})} \sum_{n_2=0}^{M_{\text{st}}^{(n_s)}(f_{c,m_{\text{ch}}})} \frac{2}{n_1!n_2!}\left[\frac{2\alpha_{1,\text{st}}^{(n_s)}(f_{c,m_{\text{ch}}})}{\sigma_{\text{st}}^{(n_s)}(f_{c,m_{\text{ch}}})}\right]^{n_1+n_2} \quad (34)$$

$$\cdot \frac{1}{4\alpha_{0,\text{st}}^{(n_s)}(f_{c,m_{\text{ch}}}) + (n_1+n_2)\sigma_{\text{st}}^{(n_s)}(f_{c,m_{\text{ch}}})}$$

$$\cdot \left\{\text{asinh}\left(\frac{\pi^2 \beta_{2,\text{eff}}^{(n_s)}(f_{c,m_{\text{ch}}}, f_{c,\text{CUT}}) \cdot \text{BW}_{\text{CUT}} \cdot (f_{e,m_{\text{ch}}} - f_{c,\text{CUT}})}{2\alpha_{0,\text{st}}^{(n_s)}(f_{c,m_{\text{ch}}}) + n_1\sigma_{\text{st}}^{(n_s)}(f_{c,m_{\text{ch}}})}\right)\right.$$

$$\left. -\text{asinh}\left(\frac{\pi^2 \beta_{2,\text{eff}}^{(n_s)}(f_{c,m_{\text{ch}}}, f_{c,\text{CUT}}) \cdot \text{BW}_{\text{CUT}} \cdot (f_{s,m_{\text{ch}}} - f_{c,\text{CUT}})}{2\alpha_{0,\text{st}}^{(n_s)}(f_{c,m_{\text{ch}}}) + n_1\sigma_{\text{st}}^{(n_s)}(f_{c,m_{\text{ch}}})}\right)\right\}$$

- The second contribution:

$$G_{\text{NLI,end}}(f) \approx \frac{16}{27} \sum_{n_s=1}^{N_s} \left[\gamma^{(n_s)}(f_{\text{CUT}}, f_{m_{\text{ch}}})\right]^2 \sum_{m_{\text{ch}}=1}^{N_{\text{ch}}} G_{\text{CUT}}^{(n_{s,\text{end}})} \left(G_{m_{\text{ch}}}^{(n_{s,\text{end}})}\right)^2 (2-\delta_{\text{CUT},m_{\text{ch}}})\left(\rho_{\text{ML}}^{(n_s)}\right)\Gamma_{n_{s,\text{end}}}^{N_s}(f_{\text{CUT}})$$

$$\cdot e^{\frac{-4\alpha_{1,\text{end}}^{(n_s)}(f_{c,m_{\text{ch}}})}{\sigma_{\text{end}}^{(n_s)}(f_{c,m_{\text{ch}}})}} \cdot \frac{1}{4\pi\beta_{2,\text{eff}}^{(n_s)}(f_{c,m_{\text{ch}}}, f_{c,\text{CUT}})} \sum_{n_1=0}^{M_{\text{end}}^{(n_s)}(f_{c,m_{\text{ch}}})} \sum_{n_2=0}^{M_{\text{end}}^{(n_s)}(f_{c,m_{\text{ch}}})} \frac{2}{n_1!n_2!}\left[\frac{2\alpha_{1,\text{end}}^{(n_s)}(f_{c,m_{\text{ch}}})}{\sigma_{\text{end}}^{(n_s)}(f_{c,m_{\text{ch}}})}\right]^{n_1+n_2} \quad (35)$$

$$\cdot \frac{1}{4\alpha_{0,\text{end}}^{(n_s)}(f_{c,m_{\text{ch}}}) + (n_1+n_2)\sigma_{\text{end}}^{(n_s)}(f_{c,m_{\text{ch}}})}$$

$$\cdot \left\{\text{asinh}\left(\frac{\pi^2 \beta_{2,\text{eff}}^{(n_s)}(f_{c,m_{\text{ch}}}, f_{c,\text{CUT}}) \cdot \text{BW}_{\text{CUT}} \cdot (f_{e,m_{\text{ch}}} - f_{c,\text{CUT}})}{2\alpha_{0,\text{end}}^{(n_s)}(f_{c,m_{\text{ch}}}) + n_1\sigma_{\text{end}}^{(n_s)}(f_{c,m_{\text{ch}}})}\right)\right.$$

$$\left. -\text{asinh}\left(\frac{\pi^2 \beta_{2,\text{eff}}^{(n_s)}(f_{c,m_{\text{ch}}}, f_{c,\text{CUT}}) \cdot \text{BW}_{\text{CUT}} \cdot (f_{s,m_{\text{ch}}} - f_{c,\text{CUT}})}{2\alpha_{0,\text{end}}^{(n_s)}(f_{c,m_{\text{ch}}}) + n_1\sigma_{\text{end}}^{(n_s)}(f_{c,m_{\text{ch}}})}\right)\right\}$$

Finally, putting the two contributions together gives the NLI power at the end of the link,

$$P^{\text{NLI}} = \frac{16}{27} \sum_{\substack{1\le m\le N_{\text{ch}}, 0\le j\le 1 \\ 0\le k\le Q_i, 0\le q\le Q_i \\ 0\le n\le N_s, 1\le i\le 2}} \frac{\gamma_{n,m}^2 \cdot P_{n,i} \cdot P_{m,i}^2 \cdot \rho_m \cdot \Gamma_{n,i}^{N_s} \cdot (2-\delta_{m,n}) \cdot e^{-4\alpha_{1,m,i}/\sigma_{m,i}} \cdot (-1)^j}{2\pi R_m^2 \cdot k!q! \cdot \bar{\beta}_{2,m} \cdot (4\alpha_{0,m,i} + (k+q)\sigma_{m,i})} \cdot \left(\frac{2\alpha_{1,m,i}}{\sigma_{m,i}}\right)^{k+q} \Psi_{m,n,j,k,i} \quad (36)$$

Where $i$ denotes the two contributions, $i=1$ is for loss profile in the forward direction, with z=0 at the span start; $i=2$ for loss profile in the backward direction, with z=0 at the span end.

$$\Psi_{m,n,j,k,i} = \text{asinh}\left(\frac{\pi^2 \bar{\beta}_{2,m} R_n \cdot (f_m - f_n + (-1)^j \cdot R_m/2)}{2\alpha_{0,m,i} + k\sigma_{m,i}}\right), Q_i = \max\left(\lfloor 10 \cdot |2\alpha_{1,n,i}/\sigma_{n,i}|\rfloor + 1\right) \quad (37)$$

## 4. Conclusion

In this paper, we provide a comprehensive derivation of CFM6, capable of evaluating the NLI induced by arbitrary Raman amplification in UWB optical systems. CFM6 builds upon CFM5 by introducing a novel contribution for calculating the NLI generated by the backward Raman amplification.